\documentclass[apj,iop]{emulateapj}
\usepackage{apjfonts}
\usepackage{amsmath}
\usepackage{natbib}
\bibpunct{(}{)}{;}{a}{}{,}
\DeclareMathAlphabet{\mathsc}{OT1}{cmr}{m}{sc}
\def\testbx{bx}%
\DeclareRobustCommand{\ion}[2]{%
\relax\ifmmode
\ifx\testbx\f@series
{\mathbf{#1\,\mathsc{#2}}}\else
{\mathrm{#1\,\mathsc{#2}}}\fi
\else\textup{#1\,{\mdseries\textsc{#2}}}%
\fi}

\slugcomment{Submitted to the Astrophysical Journal}

\shorttitle{The UV background photoionization rate from SDSS}
\shortauthors{Dall'Aglio et al.}

\begin{document}

\title{The UV background photoionization rate at $2.3\le z\le 4.6$\\ as measured from the Sloan Digital Sky Survey}

\author{ALDO DALL'AGLIO\altaffilmark{2}, LUTZ WISOTZKI\altaffilmark{2}, and G\'{A}BOR WORSECK\altaffilmark{3}}

\altaffiltext{1}{Based on observations obtained with the Sloan Digital Sky Survey, which is owned and operated by the Astrophysical Research Consortium.}
\altaffiltext{2}{Astrophysikalisches Institut Potsdam, An der Sternwarte 16, D-14482 Potsdam, Germany\\{\tt adaglio@aip.de}}

\altaffiltext{3}{UCO/Lick Observatory; University of California, Santa Cruz, Santa Cruz, CA 95064}

\begin{abstract}
We present the results from the largest investigation to date of the proximity effect in the \ion{H}{i} Ly$\alpha$ forest, using the fifth Sloan Digital Sky Survey (SDSS) data release. The sample consists of 1733 quasars at redshifts $z\gtrsim 2.3$ showing a signal-to-noise of at least $S/N=10$. We adopted the flux transmission statistic to infer the evolution of the \ion{H}{i} effective optical depth in the Lyman forest between $2\la z\la 4.5$, finding very good agreement with measurements from high-resolution quasar samples. By fitting a quasar continuum to each individual object we estimated a \emph{continuum} composite spectrum in the rest frame wavelength range $1000 \la \lambda\la 3000$~\AA. The flux in our composite spectrum closely follows a power law in the form \mbox{$f_\nu\propto\nu^{\alpha_\nu}$} with $\alpha_\nu=-0.57$ in good agreement with similar estimates inferred from significantly smaller SDSS quasar samples. We compared the average opacity close to the quasar emission with its expected behavior in the Ly$\alpha$ forest and estimated the signature of the proximity effect towards individual objects at high significance in about 98\% of the quasars. Dividing the whole sample of objects in eight subsets according to their emission redshift, we quantify the strength of the proximity effect adopting a fiducial value of the cosmic UV background photoionization rate, multiplied by a free parameter allowed to vary for different quasars. We inferred the proximity effect strength distribution (PESD) on each of the subsets finding in all cases a prominent peak and an extending tail towards values associated to a weak effect. We provide for the first time observational evidence for an evolution in the asymmetry of the PESD with redshift. Adopting the modal values of the PESDs as our best and unbiased estimates of the UV background photoionization rate ($\Gamma_\ion{H}{i}$), we determine the evolution of $\Gamma_\ion{H}{i}$ within the redshift range $2.3\le z\le 4.6$. Our measurements of UV background photoionization rate do not show any significant decline towards high redshift as expected by theoretical model predictions and are located at \mbox{$\log \Gamma_{\ion{H}{i}} = -11.78\pm 0.07$} in units of s$^{-1}$. We employ these estimates and decompose the observed photoionization rate into two major contributors: quasar and star-forming galaxies. By modeling the quasar contribution with different luminosity functions we estimated their contribution to the cosmic $\Gamma_\ion{H}{i}$, allowing us to put a constraint on the remaining amount of ionizing photons necessary to obtain the observed photoionization rate. We conclude that independently of the assumed luminosity function, stars are dominating the UV background $\Gamma_\ion{H}{i}$ at redshift $z>3$.
\end{abstract} 



\begin{figure*}
\resizebox{\hsize}{!}{\includegraphics*[]{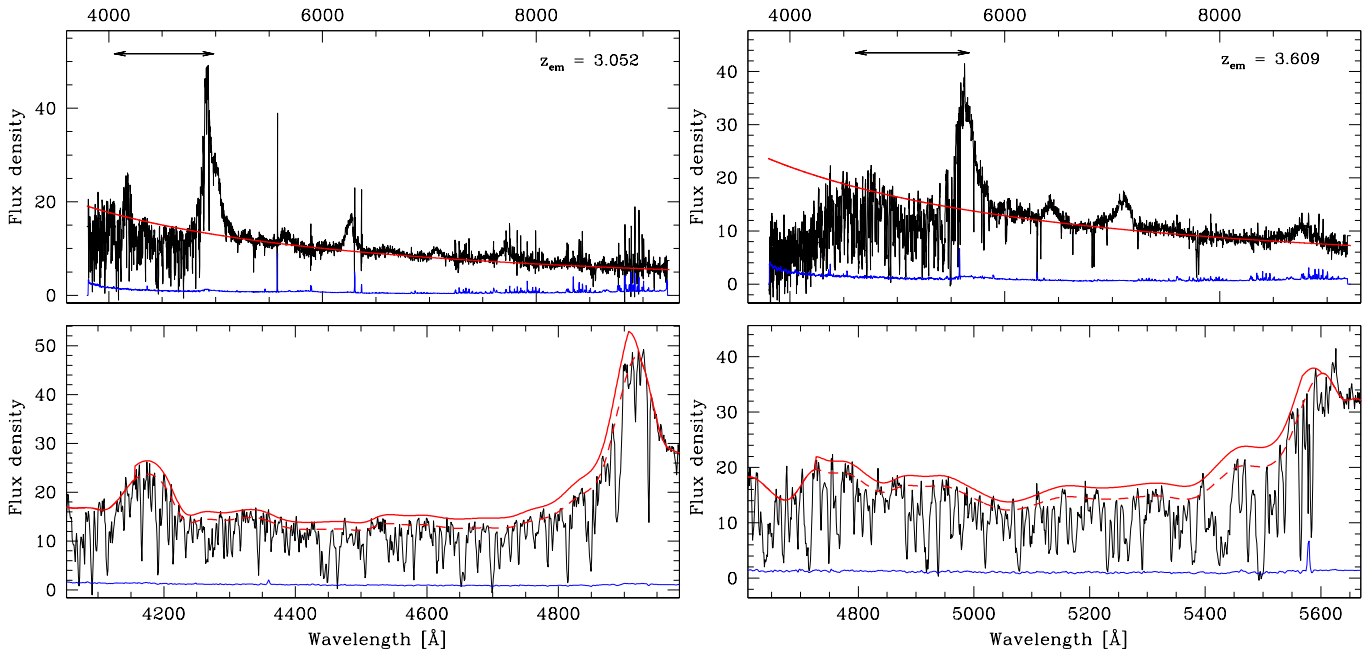}}
\caption{Two examples of quasar spectra taken from the data set with $S/N\simeq18$ (right-hand panels) and $S/N\simeq10$ (left-hand panels). In the two upper panels the spectra are plotted along with the noise arrays and the power law fit continua. The horizontal arrows over the Ly$\alpha$ forest identify the wavelength range covered in the magnified lower panels where details about the local continua are presented. In all cases the flux density is expressed in units of $10^{-17}$~erg~s$^{-1}$~cm$^{-2}$~\AA$^{-1}$. The dashed lines indicate the fitted continua while the solid lines show the corrected shape as discussed in detail in Sect.~\ref{pe_sdss_txt:continuum}.}
\label{pe_sdss_fig:qsoex}
\end{figure*}

\section{Introduction}

The intergalactic medium (IGM) is permeated by a quasi-continuous, spatially fluctuating density field, whose major constituents are hydrogen and helium. The distribution of these baryons gives rise to a multitude of narrow absorption features observed in the spectra of high redshift sources such as luminous quasars (QSOs), commonly known as the Ly$\alpha$ forest \citep{rauch98,meiksin07}. In the framework of hierarchical structure formation the IGM fills the low density regions between collapsed objects and it shows an extremely high ionization state. The lack of Ly$\alpha$ absorption troughs indicates an inhomogeneous distribution of hydrogen and a high ionization state of the IGM \citep{Gunn65}, probed so far out to $z\sim6$ \citep{fan02}. The cosmic UV background (UVB) maintaining this high ionization state is generated by the overall population of quasars and star-forming galaxies \citep{haardt96,fardal98,haardt01}. The radiation emitted by the UV sources is filtered by the IGM and diluted by cosmological expansion, leading to a redshift and frequency dependence of the UVB intensity \citep[e.g.][]{dave99,madau99,schirber03}.

Three methods are commonly employed to estimate the UVB intensity or the UVB hydrogen photoionization rate at different epochs. In the first approach, the overall luminosity density is predicted from a given UV source luminosity function, taking into account absorption and re-emission by the IGM \citep{madau99,haardt01,schirber03}. While these computations rely on challenging observations such as the galaxy luminosity function at $z>3$, they provide separate estimates on the relative contributions of quasars and star-forming galaxies to the UVB.

Secondly, the hydrogen photoionization rate can be inferred by matching the level of Ly$\alpha$ forest absorption obtained in numerical simulations to observations \citep{rauch97,theuns98,bolton05,giguere08b}. Although the physics of the IGM is well modeled in recent numerical simulations, the primary uncertainty in determining the UVB via this method is its degeneracy with respect to the observationally poorly constrained thermal state of the IGM.

The arguably most direct observational technique to infer the UVB photoionization rate relies on the proximity effect of quasars. While fluctuations in the amplitude of the UVB in the general IGM are small \citep[e.g.][]{meiksin04}, the UV radiation field near luminous quasars is expected to be enhanced. The quasar is capable of over-ionizing the gas in its vicinity, resulting in a systematic weakening if the Lyman alpha absorption within several Mpc from the quasar \citep{weymann81,carswell82,murdoch86}. The amplitude of this deficit depends on the relative contributions of ionizing photons from the local source and the UVB. Knowing the QSO luminosity, the UV background photoionization rate can be estimated \citep{carswell87,bajtlik88}.

\begin{figure*}
\resizebox{\hsize}{!}{\includegraphics*[]{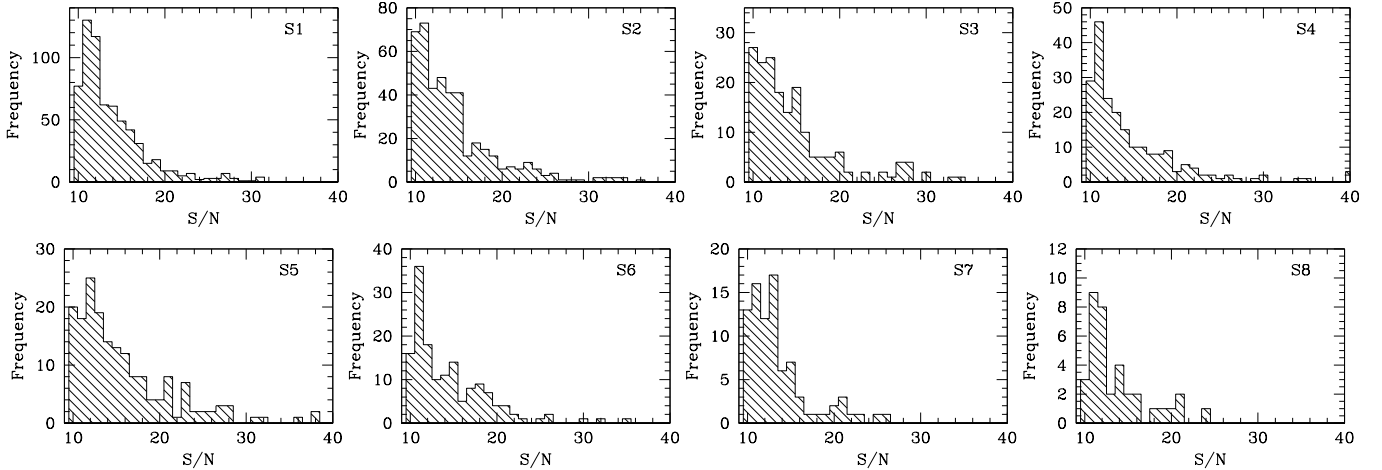}}
\caption{Signal-to-noise distribution for each of the quasar subsets as defined in Table~\ref{pe_sdss_tab:data_sample}.}
\label{pe_sdss_fig:SN_distr}
\end{figure*}

Most previous analyses showed a large scatter in the inferred UVB level  \citep[e.g.][]{giallongo96,cooke97,scott00,liske01}, owing to the relatively small sample sizes investigated. Although the proximity effect is detectable in individual QSO spectra \citep[e.g.][]{williger94,cristiani95,lu96,savaglio97,adaglio08},  the finite number of absorbers in each line of sight gives rise to a substantial random scatter that can only be compensated with sufficiently large samples \citep{adaglio08b}.

One of the key questions regarding the UV background is its evolution towards high redshift. Observations of the proximity effect of a few individual QSOs suggest a mild decline towards $z\ga 4$ \citep{williger94,lu96,savaglio97},
in broad agreement with predictions from UVB synthesis \citep{haardt01}. A similar result was obtained by us recently in an investigation of several high-resolution, high S/N quasar spectra \citep[][hereafter Paper~II]{adaglio08b}. On the other hand, two recent studies constraining the UVB from matching simulations to the observed Ly forest opacity gave a hydrogen photoionization rate nearly constant within $2 \la z \la 4$, albeit on rather different levels \citep{bolton05,giguere08b}. It is thus clear that the uncertainties are still substantial.

In the present study we build on our results obtained in Paper~II where we demonstrated the possibility to use the \emph{distribution} of proximity effect strengths measured in individual quasar spectra for constraining the UV background. Our new sample of nearly 2000 QSOs from the Sloan Digital Sky Survey is by far the largest dataset ever submitted to an analysis of the proximity effect, including a significant number of QSOs at redshifts $z\ga 4$.

The plan of the paper is as follows. We begin with a description of the spectroscopic sample of quasars in Section.~\ref{pe_sdss_txt:sample}. We then briefly describe in Section~\ref{pe_sdss_txt:simul} the code adopted to generate synthetic spectra. In Section.~\ref{pe_sdss_txt:comp_spec} and \ref{pe_sdss_txt:tau_evol} we determine the quasar composite spectrum and the evolution of the Ly$\alpha$ effective optical depth, respectively. Section~\ref{pe_sdss_txt:pe_measurements} introduces the theoretical background to measure the proximity effect and presents our results on the proximity effect signature towards individual objects. In Section~\ref{pe_sdss_txt:uvb_evol} we determine the redshift evolution of the UVB, followed by a new estimation of the relative share of quasars and star-forming galaxies to the UVB. We end with our conclusions in Section~\ref{pe_sdss_txt:concl}.  

Throughout this Chapter we assume a flat universe with Hubble constant $H_0=70$~$\mathrm{km}\,\mathrm{s}^{-1}\,\mathrm{Mpc}^{-1}$ and density parameters $\left(\Omega_\mathrm{m},\Omega_\Lambda\right)=\left(0.3,0.7\right)$.

\begin{table}
\small\centering
\caption[]{Subset definitions of the SDSS quasars.}
\label{pe_sdss_tab:data_sample}
\begin{tabular}{cccc}
\hline\hline\noalign{\smallskip}
Subset  & $z_\mathrm{q}$ & $z_\mathrm{mean}$ & Number of quasars \\
\noalign{\smallskip}\hline\noalign{\smallskip}
 S1   &  2.30 $-$  2.50  & 2.40  &   604\\
 S2   &  2.50 $-$  2.75  & 2.60  &   387\\
 S3   &  2.75 $-$  3.00  & 2.88  &   153\\
 S4   &  3.00 $-$  3.25  & 3.15  &   180\\
 S5   &  3.25 $-$  3.50  & 3.36  &   160\\
 S6   &  3.50 $-$  3.75  & 3.62  &   137\\
 S7   &  3.75 $-$  4.00  & 3.86  &   78 \\
 S8   &  4.00 $-$  4.61$^\dag$  & 4.21  &   34 \\           
\noalign{\smallskip}\hline\noalign{\smallskip} 
\end{tabular}     
\begin{list}{}{}
\item[$\dag$:] The upper limit $z_\mathrm{q}=4.61$ identifies the highest redshift object in this subset
\end{list}\end{table}       

\section{The Quasar Sample}\label{pe_sdss_txt:sample}

\subsection{Selection}\label{pe_sdss_txt:sel}

The sample of quasar spectra was drawn from the fifth data release (DR5) of the Sloan Digital Sky Survey (SDSS) quasar catalog \citep{schneider07}, with all data reprocessed by the DR6 reduction pipeline. 

The two SDSS double channel spectrographs give a resolution of $\mathrm{FWHM}\approx 2$~\AA\ ($R\sim2000$) in the wavelength range from $\sim3800$~\AA\ to $\sim9200$~\AA. We required a coverage of the Lyman alpha forest of at least $200$~\AA, corresponding to a minimum emission redshift $z_\mathrm{q}>2.3$. Thereby we were left with $\sim18\,000$ of the $\sim77\,400$ visually classified quasars in the \citet{schneider07} catalog. In order to be able to reliably measure the proximity effect in the SDSS spectra, we further imposed a minimum signal-to-noise ratio (S/N) of 10 in the quasar continuum near $1450$~\AA\ rest frame, reducing the number of QSOs to 1916. Among these there are 144 broad absorption line quasars (BAL QSOs) which cannot be used for the proximity effect analysis. Details about their identification are given in Sect.~\ref{pe_sdss_txt:bals}. Finally we inspected the remaining quasar spectra visually and removed 39 objects that were misclassified or were still suspected to be BAL QSOs. This procedure left us with a total of 1733 quasars.

The proximity effect analysis also requires accurate quasar redshifts and spectrophotometry. The redshifts in the \citet{schneider07} catalog were measured from broad emission lines in the accessible wavelength range and statistically corrected for the systematic velocity shifts between low- and high-ionization lines \citep[e.g.][]{gaskell82}. The statistical correction was determined by \citet{shen07} from a quasar sample at \mbox{$1.8<z_\mathrm{q}<2.2$}, where the SDSS spectrograph covers the \ion{Si}{iv},\ion{C}{iv}, \ion{C}{iii]}, and \ion{Mg}{ii} emission lines. Adopting the \ion{Mg}{ii} emission redshift as systemic and applying their velocity correction, the redshift accuracy of higher redshift quasars is still $\sigma\, z_\mathrm{q}\simeq 0.002$, sufficient for our purposes. Spectrophotometric accuracy is achieved by observing calibration stars together with the SDSS survey targets and tying the spectra to the accurate SDSS photometry  \citep[$\sigma_\mathrm{mag}=0.04$;][]{york00}.

In the following we divide the sample into 8 subsets according to the quasar emission redshift as listed in Table~\ref{pe_sdss_tab:data_sample}. Each subset is characterized not only by the range in redshifts, but also its number-weighted mean $z_\mathrm{mean}$. As an example of the average data quality, we show in Figure~\ref{pe_sdss_fig:qsoex} two quasar spectra together with their $1\sigma$ noise arrays and the estimated continua which will be discussed in the next section. Additionally, Fig.~\ref{pe_sdss_fig:SN_distr} shows the S/N distribution for the subsets.

\subsection{Continuum estimation}\label{pe_sdss_txt:continuum}

In order to evaluate the proximity effect along each sight line, the quasar flux at the Lyman limit and the shape of its continuum emission are needed. Therefore two types of continua were estimated: (i) a global power law \mbox{($f_\nu(\nu) \propto \nu^\alpha$)}, excluding emission and absorption regions, used to estimate the quasar flux at the Lyman limit and (ii) a more local estimate that also includes the broad emission lines as a quasi-continuum, used to infer the continuum-normalized absorption-line spectrum.

The power law continuum was fitted though pre-defined continuum regions redward of the Ly$\alpha$ emission line free of any known emission lines (Sect.~\ref{pe_sdss_txt:comp_spec}). Extrapolating the power law to the Lyman limit, we estimated the quasar Lyman limit flux. The necessity for this extrapolation is motivated by the difficulty to infer the Lyman limit flux due to the low S/N at short wavelengths or the lack of wavelength coverage at low redshifts.

\begin{figure}
\resizebox{\hsize}{!}{\includegraphics*[]{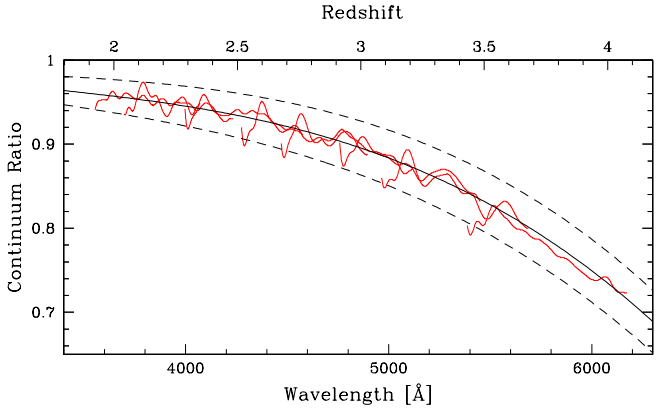}}
\caption{Average ratios between the fitted and input continua in the Ly$\alpha$ forest of our simulations (red lines). The smooth profile with the dashed lines represents the least-square second order polynomial fit to the ratios and the $\pm 1\,\sigma$ errors.}
\label{pe_sdss_fig:cont_corr_tauz}
\end{figure}

The local continuum was automatically fitted using an algorithm based on the work by \citet{young79} and \citet{carswell82}. A cubic spline was interpolated on adaptive intervals along the spectrum with respect to the continuum slope. The points for the spline interpolation were chosen starting from a regular sampling of the spectrum with a binning that becomes finer whenever the slope of the computed continuum exceeds a given threshold. This was done in order to better reproduce the wings of emission lines. To this continuum we additionally applied a systematic correction which accounts for the resolution effects and line blending, estimated from Monte-Carlo simulated spectra (see Sect.~\ref{pe_sdss_txt:simul} for details). We basically followed the procedure outlined in Paper~II and \citet{adaglio08} hereafter Paper~I. This approach consists of analyzing the ratio between the input and fitted continuum in a set of 500 mock spectra. From this ratio we determined an average continuum ratio and a dispersion used as systematic and statistical uncertainties. Our results are presented in Fig.~\ref{pe_sdss_fig:cont_corr_tauz}. The local, corrected continuum was finally used to estimate the quasar transmission $T=F_\mathrm{qso}/F_\mathrm{cont}$, defined as the ratio of the quasar flux ($F_\mathrm{qso}$) and the local continuum ($F_\mathrm{cont}$).

\subsection{Broad absorption line quasars}\label{pe_sdss_txt:bals}

The input sample of QSOs contains a significant fraction of BAL quasars which are not suitable to study the proximity effect and therefore need to be removed. To identify these objects we followed the procedure outlined by \citet{weymann91}. The approach consists of computing the balnicity index defined as
\begin{equation}
BI = -\int_{v_1}^{v_2} \left[1-T(v)/0.9\right]C \: \mathrm{d}v\label{pe_sdss_eq:bal}
\end{equation}
with $v_1 = 25000$ km\ s$^{-1}$ and $v_2 = 3000$ km\ s$^{-1}$. $T(v)$ is the transmission as function of velocity and $C$ is a constant initially set to zero. $C$ becomes 1.0 whenever the term in the brackets remains positive for at least 2000 km\ s$^{-1}$, meaning that the first 2000 km\ s$^{-1}$ of a BAL do not contribute to the balnicity index. $C$ returns to zero whenever the quantity in brackets becomes negative. If a quasar shows a positive balnicity index, it was flagged and removed from our sample. 

We note that in spite of the seemingly objective definition of the balnicity index, the identification of a BAL quasar still depends on the subjective choice of a number of factors such as the quasar continuum or the velocities $v_1$ and $v_2$. With this procedure we identified 144 BAL quasars not suitable for the analysis of the proximity effect. We noticed later that a few objects with BAL-like features survived this deselection and remained in the sample. We discuss their role in the proximity effect detection in Sect.\ref{pe_sdss_txt:pepd}.

\section{Monte Carlo simulations}\label{pe_sdss_txt:simul}

Simulating quasar spectra provides a powerful tool of comparison which we employ to accurately control and calibrate a number of systematic effects present in real observations. We employed such simulations for three main purposes. (i) We estimated the systematic and statistical uncertainties of our continuum placement in the quasar spectra (Sect.~\ref{pe_sdss_txt:continuum}). (ii) These uncertainties were used to obtain error bars on the effective optical depth in the forest, in addition to the variations in Ly$\alpha$ absorption among different lines of sight  (Sect.~\ref{pe_sdss_txt:tau_evol}). (iii) We used the simulated spectra to assess the bias arising from an asymmetric proximity effect strength distribution when combining the data to estimate the UVB photoionization rate (Sect.~\ref{pe_sdss_txt:uvb_evol}).

The procedure used to generate synthetic spectra is identical to the one presented in Paper~II. We generated a set of 500 synthetic spectra adopting a simple Monte Carlo prescription, assuming that the Ly$\alpha$ forest is well represented by three observed distributions:
\begin{enumerate}
\item{The line number density distribution, approximated by a power law of the form \mbox{$\mathrm{d}n / \mathrm{d}z \propto (1+z)^{\gamma}$} where \mbox{$\gamma=2.04$} (Paper~II).}
\item{The column density distribution, given by \mbox{$f(N_{\mbox{\scriptsize\ion{H}{i}}}) \propto N_{\mbox{\scriptsize\ion{H}{i}}}^{-\beta}$} where the slope is \mbox{$\beta \simeq 1.5$} \citep{kim01}.}
\item{The Doppler parameter distribution, given by  \mbox{$\mathrm{d}n / \mathrm{d}b \propto b^{-5}\mathrm{exp}\left[{-{b_{\sigma}^4}/{b^4}}\right]$} where  \mbox{$b_{\sigma}\simeq 24\;\mathrm{km/s}$} \citep{kim01}.}
\end{enumerate}
The algorithm populated the simulated sight lines with absorption features following the above distributions until the evolution of the effective optical depth matches the estimates made in Paper~II. Once the simulated transmission was computed, an artificial quasar spectral energy distribution including emission lines of varying strengths and widths was generated via the principal component method described by \citet{suzuki06} and in Paper~I. The resolution of the spectrum was degraded using a Gaussian smoothing function with a FWHM of approximately 2\AA\ and pixel size increased to match the SDSS specifications. Gaussian noise was added to the final quasar spectrum, in order to reproduce the S/N level of our objects.

\begin{figure}
\resizebox{\hsize}{!}{\includegraphics*[]{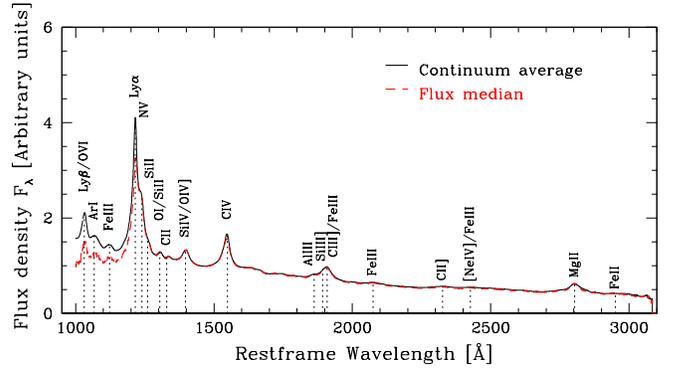}}
\caption{The flux-median and continuum-mean SDSS composite spectra derived from our set of QSOs. The vertical dotted lines mark the positions of the most prominent emission lines known in quasar spectra. The difference between the two composites blueward of the Ly$\alpha$ emission line is due to the influence of the Ly$\alpha$ forest absorption.}
\label{pe_sdss_fig:comp_sper_lab}
\end{figure}
\begin{figure}
\resizebox{\hsize}{!}{\includegraphics*[]{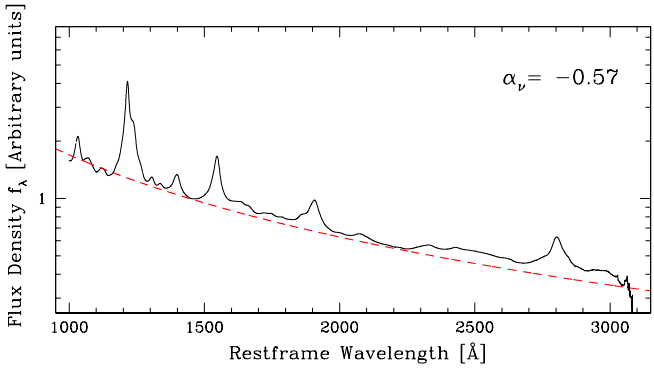}}
\caption{The dashed line represents the power law fit ($f_\nu(\nu) \propto \nu^\alpha$) to the \emph{continuum-mean} SDSS composite spectrum.}
\label{pe_sdss_fig:comp_spec}
\end{figure}

\section{A quasar composite spectrum}\label{pe_sdss_txt:comp_spec}

The large number of objects in our sample motivated us to compute a composite quasar spectrum. This spectrum provides important insights about the position of low and high-ionization lines as well as global properties of the quasar spectral energy distribution. The computation of the composite spectrum required: (i) A rebinning of the spectra in rest frame wavelengths, (ii) a normalization of the fluxes and (iii) a combining method to obtain the final composite spectrum. We computed two types of composite spectra, each of which has a different combining procedure: (1) the median spectrum employing the fluxes of the QSOs to minimize the effect of absorption lines and (2) the mean spectrum adopting the fitted continua of the QSOs. In the latter approach all quasar continua were corrected for the systematic depression in the Ly$\alpha$ forest as described in Sect.~\ref{pe_sdss_txt:continuum}. Both results are presented in Fig.~\ref{pe_sdss_fig:comp_sper_lab}. The two composites are essentially identical until the Ly$\alpha$ emission where the differences between the two become visible. These differences are due to the absorption in the Ly forest which we accounted for when combining the corrected quasar continua. 

The very high S/N level of our continuum composite spectrum allowed us to identify many of the well known emission lines present in the spectra of high redshift QSOs. To confirm the accuracy of the quasar redshifts provided in the \citet{schneider07} catalog, we measured from the composite spectrum the location of the most prominent low and high-ionization emission lines. For lines such as the Ly$\alpha$, \ion{Si}{ii}+\ion{O}{i}, \ion{Si}{iv}+\ion{O}{iv]}, \ion{C}{iv}, \ion{C}{iii}, and \ion{Mg}{ii} we fitted one or more Gaussian profiles to accurately estimate the emission wavelengths. The measured values for low-ionization lines were in agreement with their laboratory wavelengths while the high-ionization lines presented a systematic blueshift up to 550 km/s in the case of \ion{C}{iv} \citep{gaskell82,tytler92,richards06}.

The computation of the two composite spectra enabled us to check the accuracy of the BAL quasars removal. If a significant fraction of such objects were misidentified, we should have detected a depression only in the flux composite, redward of the \ion{C}{iv} emission line. The lack of such a feature ensures a correct removal of any significant contamination by BAL quasars.

The mean continuum composite spectrum is plotted in Fig.~\ref{pe_sdss_fig:comp_spec} with a single power law fit. The accuracy of this fit is compromised by the limited number of regions in a QSO spectrum free of emission lines. To perform the power law fit we adopted the three regions between 1350$-$1365~\AA\, 1445$-$1475~\AA\ and 2200$-$2250~\AA. The estimated spectral slope is $\alpha_\nu=-0.57$, in agreement with previous measurements based on earlier data releases \citep{vandenberk01}. This value is further supported by the distribution of spectral indices which closely follows a Gaussian distribution centered at $\alpha_\nu=-0.6$ with a dispersion of $\sigma_\alpha=0.35$. Therefore, estimating the quasar flux at the Lyman limit via a single power law extrapolation ensures reasonable accuracy in the computation of the proximity effect.

\begin{figure*}
\centering
\includegraphics*[width=12cm]{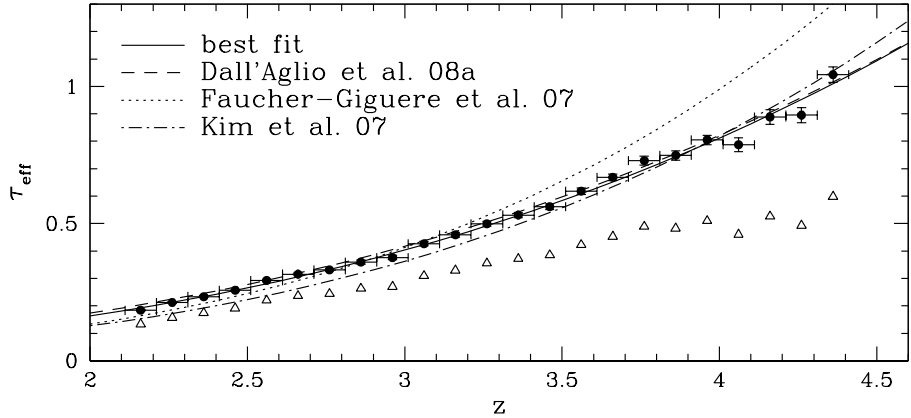}
\caption{Observed effective optical depth as a function of redshift with and without the continuum correction (dots and triangles, respectively). The solid profile shows the power-law least-squares fits to the data points while the dashed, dotted and dashed-dotted curve indicates the evolution of the Ly$\alpha$ optical depth measured in Paper~II and, by \citet{giguere07} and \citet{kim07} respectively, using high resolution spectra. Errors on the redshift axis indicate the portion of forest used to compute the effective optical depth. The uncertainties on the effective optical depth have been estimated using our Monte Carlo simulated sight lines. }
\label{pe_sdss_fig:tau_evol}
\end{figure*}

\begin{table}
\centering
\caption[]{Effective optical depth in the Ly$\alpha$ forest of our combined QSO sample, corrected for the continuum uncertainties.}
\label{pe_sdss_tab:tau_comb}
\begin{tabular}{cccccc}
\hline\hline\noalign{\smallskip}
$<z>$ & $\tau_\mathrm{eff}$ & $\sigma(\tau_\mathrm{eff})$ & $<z>$ & $\tau_\mathrm{eff}$ & $\sigma(\tau_\mathrm{eff})$\\
\noalign{\smallskip}\hline\noalign{\smallskip}
 2.16 & 0.183  & 0.002  &  3.36 & 0.531  & 0.009\\
 2.26 & 0.211  & 0.003  &  3.46 & 0.561  & 0.010\\
 2.36 & 0.233  & 0.003  &  3.56 & 0.617  & 0.011\\
 2.46 & 0.256  & 0.003  &  3.66 & 0.668  & 0.012\\
 2.56 & 0.291  & 0.004  &  3.76 & 0.729  & 0.016\\
 2.66 & 0.315  & 0.004  &  3.86 & 0.748  & 0.016\\
 2.76 & 0.331  & 0.004  &  3.96 & 0.804  & 0.017\\
 2.86 & 0.359  & 0.008  &  4.06 & 0.787  & 0.025\\
 2.96 & 0.376  & 0.008  &  4.16 & 0.888  & 0.026\\
 3.06 & 0.426  & 0.008  &  4.26 & 0.895  & 0.027\\
 3.16 & 0.459  & 0.008  &  4.36 & 1.043  & 0.028\\
 3.26 & 0.499  & 0.008  & &&\\
\noalign{\smallskip}\hline\noalign{\smallskip} 
\end{tabular}     
\end{table}       

\section{Effective optical depth of the Ly$\alpha$ forest}\label{pe_sdss_txt:tau_evol}

The signature of the proximity effect is typically observed against the increase with redshift of the mean absorption in the Ly$\alpha$ forest. The opacity of the Ly$\alpha$ forest is empirically known to scale with redshift according to a power law of the form
\begin{equation}
\tau_{\rm{eff}}= \tau_0 (1+z)^{\gamma+1}.\label{pe_sdss_eq:tau}
\end{equation}
for a line density evolving as $\mathrm{d}n/\mathrm{d}z \propto (1+z)^\gamma$. In Equation~\ref{pe_sdss_eq:tau} $\tau_{\rm eff}$ is the \ion{H}{i} effective optical depth. This quantity is related to the mean transmission as $\tau_{\rm eff} \equiv -\ln\left <T\right>$, where the average $\left < \: \right>$ is taken in redshift intervals.

The adopted method to estimate the parameters $(\log\tau_0,\gamma)$ consists of measuring $\tau_{\rm eff}$ in the forest of those quasars which intersect a particular redshift slice ($\Delta z = 0.1$). We removed the influence of the quasar by considering only the rest frame wavelength range $1025<\lambda<1180$~\AA. We performed this computation first on the transmission without continuum correction, and then applied the correction as described in Sect.~\ref{pe_sdss_txt:continuum}. 

The results are shown in Fig.~\ref{pe_sdss_fig:tau_evol} and listed in Table~\ref{pe_sdss_tab:tau_comb}. It is evident that the continuum correction is crucial to prevent strong biases in measuring the evolution of $\tau_{\rm eff}$ at low resolution. The best fit of Eq.~\ref{pe_sdss_eq:tau} to the continuum-corrected data yields values of $(\log\tau_0,\gamma) = (-2.28 \pm 0.03,2.13 \pm 0.04)$. Despite the moderate spectral resolution, our results are in excellent agreement with the evolution of the Ly$\alpha$ optical depth as recently measured by us from high resolution quasar spectra in Paper~II and also by different authors \citep[e.g.][]{schaye03,kim07}.
We also observe that for $z\ga 3.5$ our estimated $\tau_{\rm{eff}}(z)$ is significantly lower than recent measurements by \citet{giguere07}, based on a sample of 84 QSO spectra obtained by different instruments. We have no explanation for this discrepancy. Because of the excellent agreement with other studies we proceed adopting our newly estimated fit parameters for further computations.

As a parenthetical note we remark that we do not detect any deviation from a single power law evolution at $z\simeq 3.2$. Around this redshift, \citet{bernardi03} detected a dip in $\tau_{\rm{eff}}(z)$ based on a much smaller sample of SDSS quasar spectra and speculated that this might be related to \ion{He}{ii} reionization. While our higher resolution data do show a similar dip (cf.\ Paper~II), there is no evidence for such a feature in Fig.~\ref{pe_sdss_fig:tau_evol}. A more detailed analysis of the reality of that dip is probably worthwhile, but outside the scope of the present paper.

\section{Measurements of the proximity effect}\label{pe_sdss_txt:pe_measurements}

\subsection{Definitions and procedure}\label{pe_sdss_txt:flux_stat}

The ionization state of the IGM near a quasar will be dominated by the local source rather than the UVB, leading to the observed weakening of the Ly$\alpha$ absorption in its vicinity. An efficient technique of detecting the proximity effect is to compare the effective optical depth measured near a quasar to its expected value in the Lyman alpha forest \citep[e.g.][]{Liske98,liske01}. The flux statistic technique is applicable to high- and low-resolution spectra and has been used by us to detect the proximity effect on individual lines of sight (Paper~I \& Paper~II). 

The influence of the quasar modifies the effective optical depth to
\begin{equation}
\tau_\mathrm{eff}= \tau_0  (1+z)^{\gamma+1}(1+\omega)^{1-\beta}\label{pe_sdss_eq:tauPE}
\end{equation}
where $\beta$ is the slope in the column density distribution. The quantity $\omega$ is defined as the ratio between the quasar and background photoionization rates: 
\begin{equation}
\omega(z)= \frac{\Gamma_\mathrm{q}\left(z\right)}{\Gamma_\mathrm{\ion{H}{i}}\left(z\right)} = \frac{1}{\Gamma_\mathrm{\ion{H}{i}}\left(z\right)}\int_{\nu_0}^{\infty}{\frac{f_{\nu}\left(\nu,z\right)}{h\nu}\,\sigma\left(\nu\right)\mathrm{d}\nu}.\label{pe_sdss_eq:omega_ratio}
\end{equation}
We assume the UVB photoionization rate to be reasonably constant for the redshift path length probed by the Ly$\alpha$ forest of a given QSO. Approximating the spectral energy distribution of a quasar in the form $f_\nu(\nu) \propto \nu^{\alpha_\mathrm{q}}$, we can rewrite $\Gamma_\mathrm{q}$ as:
\begin{equation}
\Gamma_\mathrm{q}\left(z\right)=\frac{\sigma_\nu(\nu_0)\ L_\nu(\nu_0)}{4 \pi h\ (3-\alpha_\mathrm{q})\ d^2_{L}(z_{\mathrm{q}},z)}  \left(\frac{1+z_\mathrm{q}}{1+z}\right)^{\alpha_\mathrm{q}+1}
\label{pe_sdss_eq:qso_gamma}
\end{equation}
where $z$ is the redshift along the line of sight such that $z<z_{\mathrm{q}}$, $d_{L}(z_{\mathrm{q}},z)$ is the luminosity distance of the QSO  as seen from any foreground redshift along the LOS, $\alpha_\mathrm{q}$ is the spectral slope of the QSO and $L_{\nu_0}$ its luminosity at the Lyman limit. The hydrogen photoionization cross section at the Lyman limit is $\sigma_\nu(\nu_0)=6.33\times 10^{-18}$~cm$^2$ and $h$ is the Planck's constant.

We emphasize that all previous analyses of the proximity effect were based on a simplified determination of $\omega(z)$ as a flux ratio by assuming the same spectral energy distribution of the quasar and the cosmic UV background in Equation~\ref{pe_sdss_eq:omega_ratio}. With our correct definition of omega as a ratio of photoionization rates, we do not have to assume a UVB spectral energy distribution to get poorly constrained UVB intensities. Moreover, all recent numerical studies of the UVB give the UVB photoionization rate, facilitating comparisons between different methods.

\begin{figure*}
\resizebox{\hsize}{!}{\includegraphics*[]{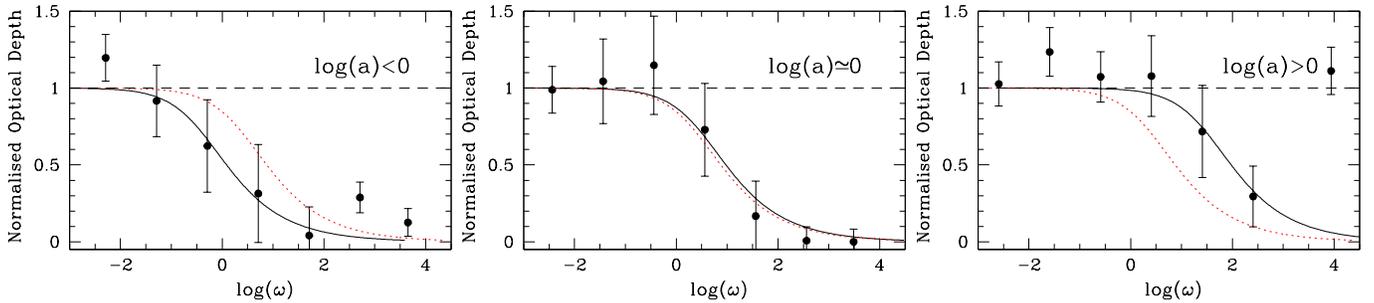}}
\caption{Three examples of the proximity effect signature along individual sight lines at redshift $z\sim3$. Each panel shows the normalized effective optical depth $\xi$ versus $\omega$, binned in steps of $\Delta\log\omega = 1$, with the fiducial model superimposed as the dotted line. The solid lines delineate the best fit to each individual QSO as described in the text. This subset was chosen for presentation purposes to show the variable strength of the proximity effect, going from \emph{strong} (left panel) to \emph{weak} (right panel).}
\label{pe_sdss_fig:pe_plot}
\end{figure*}

Finally, the ratio of the observed optical depth to the one expected in the Ly$\alpha$ forest, or the \emph{normalized effective optical depth} $\xi$, is given by
\begin{equation}
\xi=\frac{\tau_\mathrm{eff}}{\tau_0 (1+z)^{\gamma+1}}=(1+\omega)^{1-\beta}. \label{pe_sdss_eq:xi}
\end{equation}
with the parameters $(\log\tau_0,\gamma)$ measured in Sect.~\ref{pe_sdss_txt:tau_evol}. The slope of the column density distribution was fixed to $\beta = 1.5$ \citep{kim01}.

The proximity effect will manifest its signature as a departure from unity in the normalized optical depth as $\omega \rightarrow \infty$, i.e. towards the quasar systemic redshift.

\subsection{The proximity effect along single sight lines}\label{pe_sdss_txt:sglos}

We now explore the application of the above method to individual objects. Our approach is as follows: We first compute, for each individual line of sight, the normalized effective optical depth as a function of $\omega$, within bins of $\log\omega$. We then search for a systematic decrease of $\xi$ at high values of $\omega$. 

In order to compute the $\omega$ scale (Eq.~\ref{pe_sdss_eq:omega_ratio}), a fiducial value for the UVB photoionization rate had to be assumed. We fixed this value to $\Gamma^\star = 10^{-12}\,\mathrm{s}^{-1}$, thereby uniquely converting the redshift scale into an $\omega$ scale. Dividing the  $\log \omega$ axis into an equally spaced grid, we then modeled the decrease of $\xi(\omega)$ in bins of $\Delta\log\omega$ according to the formula
\begin{equation}
F(\omega)=\left(1+\frac{\omega}{a}\right)^{1-\beta}\label{pe_sdss_eq:fit}
\end{equation}
where $a$ is the only free parameter which expresses the \emph{observed} turnover of $\xi$ in units of $\Gamma^\star$. In principle, each $a$ directly provides a best-fit value of $\Gamma_{\ion{H}{i}}$, since $\Gamma_{\ion{H}{i}} = a\times \Gamma^\star$. As demonstrated in Paper~II, however, the random scatter of $a$ due to the finite small number of absorbers in each line of sight is so large that meaningful constraints on $\Gamma_{\ion{H}{i}}$ can only be obtained from the distribution of $a$ values for a statistically significant sample.

For a given sight line, the value of the fitting parameter $a$ describes how much the best-fit model disagrees with the fiducial value assumed above. We regard $\log a$ as a measure of the strength of the proximity effect, which can be classified into three principal regimes as presented in Fig.~\ref{pe_sdss_fig:pe_plot}:
\begin{enumerate}
\item $\log a < 0$: this case describes a \emph{strong} proximity effect. The normalized effective optical depth is already small for $\omega\ll 1$. 
\item $\log a \simeq 0$: this case describes an \emph{average} proximity effect since the $\xi$ values follow the fiducial model profile.
\item $\log a > 0$: this case describes a \emph{weak} proximity effect. The normalized effective optical depth will turn over only at very large $\log \omega$ values (if at all).
\end{enumerate}
Of course, the boundaries between these three categories depend on the (arbitrary) choice of the fiducial UVB photoionization rate value $\Gamma^\star$ which fixes the zero point of $\log a$.

\begin{figure*}
\resizebox{\hsize}{!}{\includegraphics*[]{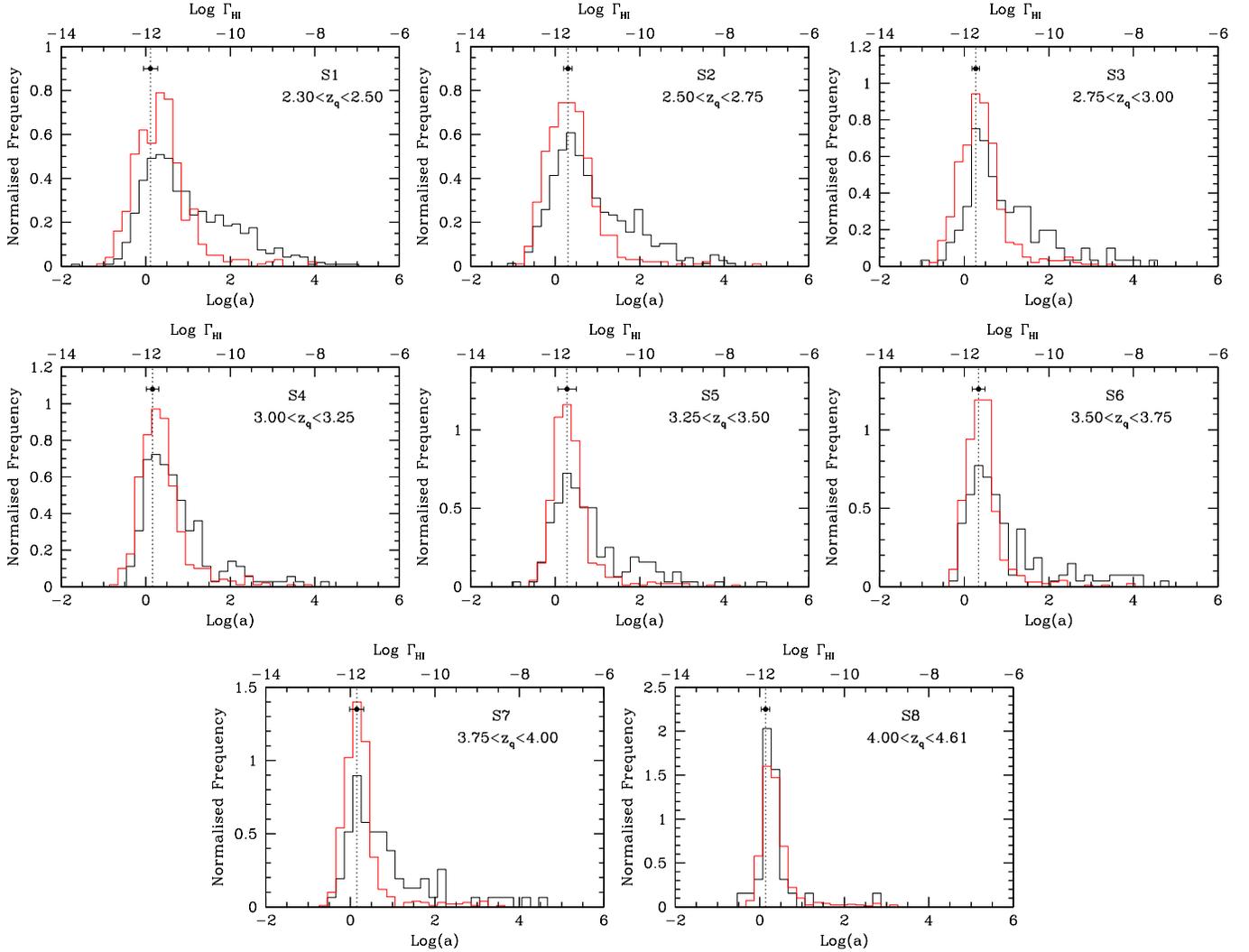}}
\caption{The proximity effect strength distributions in individual subsets of our data set. Each panel shows the observed PESD for increasing redshift intervals (black histogram) along with the simulated PESD obtained from a set of simulated lines of sight (gray histogram). The dotted line and the point mark the modal values of the PESD with the associated uncertainties.}
\label{pe_sdss_fig:log_a_distr}
\end{figure*}

\subsection{The distribution of proximity effect strengths}\label{pe_sdss_txt:pepd}

We quantified the proximity effect in the above manner on each sight line in the sample to compute the distribution of $\log a$. Considering the UVB as a constant for each subset of Table~\ref{pe_sdss_tab:data_sample} we plot the observed proximity effect strength distribution (PESD) of each subset in Fig.~\ref{pe_sdss_fig:log_a_distr}.

Each of these distributions shows a wide spread in $\log a$ values.  For in total 35 objects, or 1.8\,\% of the sample, we obtain $\log a > 3.7$, which means that for these objects the fit hinges on a single point showing a decrease of $\xi(\omega)$ at large $\omega$. Since the largest values of $\log a$ correspond to the weakest proximity effect signatures, we inspected these cases individually. It turned out that all of them displayed either strong associated absorption or were previously unrecognized BAL QSOs. For all other QSOs in the sample, the proximity effect signature is based on at least two and in general several points in the $\xi$ vs.\ $\log\omega$ diagram; we conclude that the effect is detected in 98\,\% of the sample.

For all subsamples the PESD shows the two major characteristics discovered in Paper~II: (i) A well defined peak, and (ii) a significant asymmetry with an extended tail towards weaker proximity effects, i.e.\ large values of $a$. Additionally, the PESD becomes substantially more narrow with increasing redshift and also shows a less prominent tail towards $\log a>0$.

The skewness in the PESD is the result of the $\omega$ definition as a function of redshift (Eq.~\ref{pe_sdss_eq:omega_ratio}). Approaching the quasar emission, equal $\Delta \log \omega$ bins probe progressively smaller $\Delta z$ intervals, leading to a deviation in the distribution of the absorber counts from a Gaussian. This translates to a skewed distribution of $\xi$ as $z\rightarrow z_\mathrm{q}$ since large $\xi$ values (and therefore a weaker observed proximity effect) become more likely. This trend becomes less prominent at high redshift because equal $\Delta \log \omega$ bins will probe progressively increasing numbers of absorbers and therefore reduce the skewness in the PESD. 

The PESD asymmetry is further amplified by the presence of overdensities near some of the QSOs in each subset. In Paper~II we found that about $10\%$ of our objects showed a significant excess of absorption within a few Mpc of the quasar redshift. The inferred values of $a$ for these quasars are expected to be larger by a factor related to the amount of absorption which becomes observable in the PESD as an enhancement in the tail extension on its weak side. Since we do not have sufficiently high resolution in our spectra to estimate the excess of absorption on Mpc scales near the quasar emission, we estimated its influence by employing our Monte Carlo simulation. 

We generated a set of synthetic quasar spectra as described in Sect.~\ref{pe_sdss_txt:simul}, matching the characteristics of each of our subsets (sample size, and redshift and luminosity distributions). For each spectrum we performed a continuum fit, also applying the appropriate correction and then we estimated the proximity effect strength. This exercise serves two purposes: (i) as an accuracy check for our automatic data analysis, and (ii) as a tool to reveal the effect of \ion{H}{i} overdensities on the PESD. For the input photoionization rates in the different subsets we adopted the modal value of the observed PESDs (see Sect.~\ref{pe_sdss_txt:uvb_method}).

Each panel of Figure~\ref{pe_sdss_fig:log_a_distr} shows the comparison between the observed and the simulated PESD. These distributions clearly differ with respect to the tail extension towards weaker proximity effect, while they are in agreement with respect to the location of the most likely value of $\log a$. The simulated PESD is narrower than the observed one, indicating that a certain fraction of QSOs may be affected by large-scale overdensities. Our approach to estimate the UV background from these distributions rests on the assumption that the peak of the PESD is unaffected by this contamination. We discuss the validity of this assumption further below.

\section{The evolution of the UV background photoionization rate}\label{pe_sdss_txt:uvb_evol}

\subsection{Method}\label{pe_sdss_txt:uvb_method}

As shown in Paper~II, the shape of the PESD depends on the cosmic photoionization rate as well as on possible excess \ion{H}{i} absorption on scales of several Mpc around the quasars. We now estimate the photoionization rate $\Gamma_{\ion{H}{i}}$ in various redshift bins by evaluating the PESD.

The first method consists of recovering the modal value of the PESD, i.e. estimating the peak of the distribution. To achieve this goal we adopted the following bootstrap technique: Starting from the observed distribution of $N_i$ values of $\log a$, where $N_i$ is the total number of objects in a given subset, we randomly duplicated the strength parameter of $N_i/e$ quasars and estimated the modal value of the new PESD. We repeated this process 500 times in each of the eight subsets, obtaining the mean and the sigma values of PESD modes.

\begin{figure}
\resizebox{\hsize}{!}{\includegraphics*[]{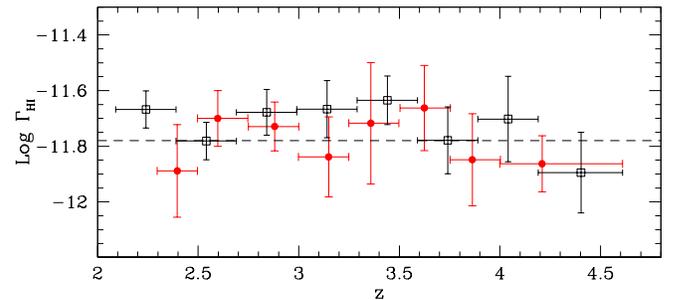}}
\caption{Comparison of the two different estimates of the UVB photoionization rate inferred from the modal value of the PESDs (solid dots) and from the combined analysis of the proximity effect after truncating the PESD as described in the text (empty squares). We also show the best-fit non-evolutionary model to our data (dashed line) yielding a photoionization rate of $\log \Gamma_{\ion{H}{i}} = -11.78\pm 0.07$ in units of s$^{-1}$.}
\label{pe_sdss_fig:uvb_evol_pesd}
\end{figure}

\begin{figure*}
\resizebox{\hsize}{!}{\includegraphics*[]{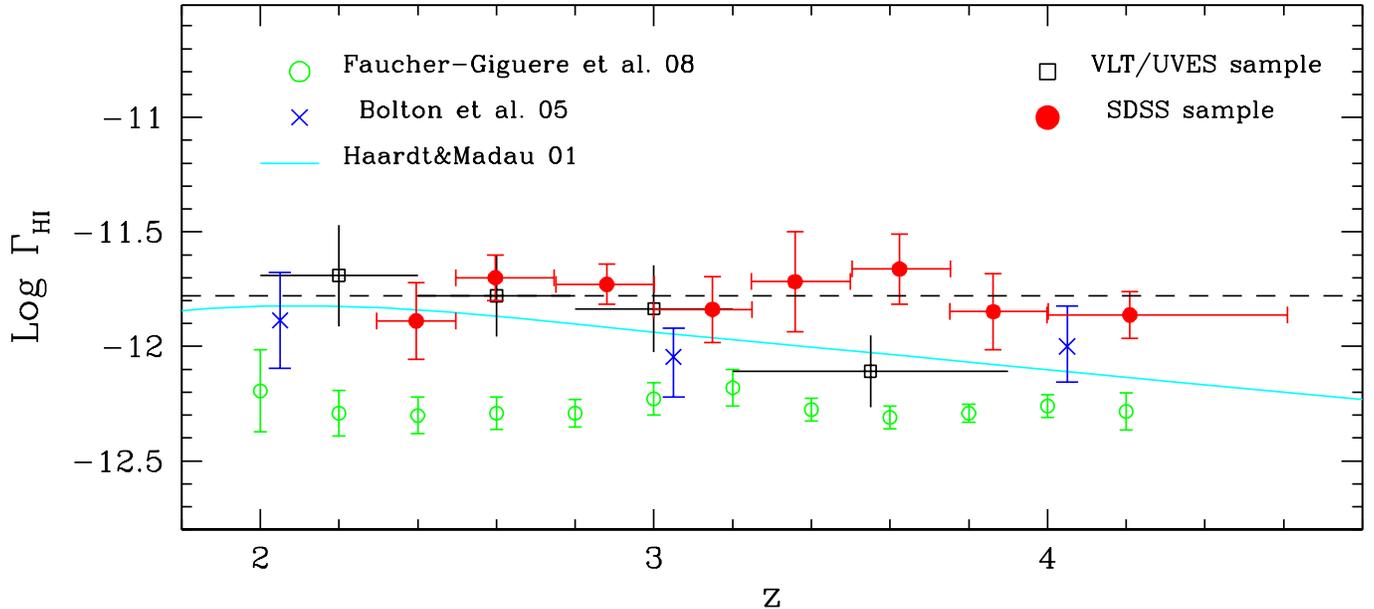}}
\caption{Evolution of the UVB  photoionization rate as obtained employing the modal value of the PESDs on the SDSS data set (solid circles). The empty squares refer to the estimates from the UVES sample presented in Paper~II. We compare our results with recent estimates from \citet{bolton05} and \citet{giguere08b}, both employing numerical simulations to infer the cosmic photoionization rate (crosses and empty circles, respectively).}
\label{pe_sdss_fig:uvb_evol_comb}
\end{figure*}

In addition to the modal values of the PESD, we also explored another method to estimate the photoionization rate evolution with redshift. We take advantage of the fact that overdense quasars will have statistically larger $\log a$ values and selected from our sample only those QSOs with $\log a$ below a certain threshold $\log a_0$. We fixed $\log a_0=1.5$ after comparing how the simulated and observed PESD change with redshift (see Fig.~\ref{pe_sdss_fig:log_a_distr}). This is equivalent to truncating the PESD above  $\log a_0$. Such a cut does not modify the estimates of the PESD mode, but significantly reduces the biases connected to overdense quasars. We then combined the signal of all these quasars by merging them according to their redshift scale. We derived the $\xi$ values in the same way as for measuring the proximity effect along single sight lines and then fitted Eq.~\ref{pe_sdss_eq:fit} to the data, yielding a first estimate of $\Gamma_{\ion{H}{i}}$. We finally corrected the preliminary photoionization rates for the systematic biases introduced by the remaining PESD asymmetry.

Figure~\ref{pe_sdss_fig:uvb_evol_pesd} shows a comparison between our two estimates of $\Gamma_{\ion{H}{i}}$. The UVB values derived via both methods are consistent with another, yielding a roughly constant level of the \ion{H}{i} photoionization rate at $2\la z\la 4.6$. A least-square fit to a non-evolving photoionization rate yields \mbox{$\log \Gamma_{\ion{H}{i}} = -11.78\pm 0.07$} for the PESD modal values and \mbox{$\log \Gamma_{\ion{H}{i}} = -11.75\pm 0.05$} from the second method, both in units of s$^{-1}$. While the two methods yield statistically very similar results, we adopted the modal estimates of the UVB photoionization rate as our final estimates. This choice is mainly motivated by the conceptual simplicity of the PESD mode and its independence of  \emph{ad hoc} assumptions regarding the threshold $\log a_0$ or the PESD asymmetry correction.

\subsection{Effect of overdensities on the proximity effect strength distribution}\label{pe_sdss_txt:over_den}

We now address the concern that large-scale \ion{H}{i} overdensities around our quasars might significantly bias the UVB estimates. A fixed overdensity factor valid for \emph{all} quasars in a sample would simply shift the PESD towards larger $\log a$ values, but preserve the shape of the distribution. Our method is completely degenerate to such a shift. We argued in Paper~II that a uniform overdensity factor is however rather unplausible. In the more likely case of 
a distribution of overdensities the width of the PESD would be significantly enhanced. Measuring the width of the distribution thus provides a useful additional diagnostic.

In Fig.~\ref{pe_sdss_fig:log_a_distr} we compare the measured PESDs with simulated distributions generated by the Monte-Carlo process as described above. In each panel, the simulated dataset has the same redshift distribution as the observed one, and the areas under the observed and simulated histograms are normalized to the same total number of objects. 

In most of the panels of Fig.~\ref{pe_sdss_fig:log_a_distr}, the two histograms agree quite well in the cores, but the observed data show a significant excess of large $\log a$ values that is absent in the simulations. In terms of standard deviations, the observed distributions have dispersions typically $\sim 1.5\times$ that of the simulated ones; most of that excess is due to the objects with $\log a \ga 2$. We interpret this excess as the main manifestation of \ion{H}{i} overdensities around our quasars.

The applicability of our approach to estimate the UVB via the proximity effect now rests on the question whether or not the mode of the PESD is significantly affected by the influence of overdensities. The generally good agreement between the \emph{cores} of the simulated and observed distributions suggests that only a minority of QSOs produce the excess; this supports our notion that the mode is approximately conserved. 

Consider our highest redshift subset S8 in particular, containing 34 QSOs at $z > 4$. The PESD of this subset has a dispersion of $\sigma_{\log a}= 0.51$, while the corresponding simulated PESD has 
$\sigma = 0.46$. Essentially all of this difference is due to a single outlier. Any excess \ion{H}{i} absorption in the QSOs would make the PESD broader and inconsistent with the observed narrow distribution, unless the amount of such an excess absorption were nearly exactly equal for the entire sample.

We conclude that the PESD mode is quite robust against the influence of overdensities. While we cannot firmly exclude the possibility that some of our $\Gamma_{\ion{H}{i}}$ values at intermediate redshifts are somewhat affected, the narrowness of the $z\ga 4$ PESD provides evidence against strong \ion{H}{i} overdensities on scales of several Mpc. We consider this data point as our most significant measurement.

\subsection{Comparison with other measurements}\label{pe_sdss_txt:dis}

Figure~\ref{pe_sdss_fig:uvb_evol_comb} shows our estimated values of the hydrogen photoionization rate $\Gamma_{\ion{H}{i}}$ as a function of redshift, together with recent results from the literature. In order to facilitate a direct comparison, we repeated the proximity effect analysis of our high-resolution VLT/UVES spectra (Paper~II), but now quoting photoionization rates rather than UVB intensities and applying a coarser binning. The redshift-averaged $\Gamma_\mathrm{\ion{H}{i}}$ values are very similar, $-11.78\pm0.07$ for SDSS vs.\ $-11.83\pm0.10$ for UVES (in logarithmic units of s$^{-1}$). While at $z\la 3$ the two datasets are almost indistinguishable, the UVES measurements suggest a decrease of $\Gamma_\mathrm{\ion{H}{i}}$ with increasing $z$ while the SDSS data appear to favor an approximately constant photoionization rate.  The differences are not highly significant considering the error bars; we also recall that the UVES sample contains only 7 QSOs with $z>3.3$. 

The SDSS measurements are also in broad agreement with the results by \citet{bolton05} who inferred $\Gamma_\mathrm{\ion{H}{i}}$ by matching hydrodynamical simulations to the observed opacity evolution in 19 QSOs (with only 4 located at $z>3.3$). Bolton et al.\ discussed the main uncertainties entering in the computation such as the IGM temperature, the IGM density distribution and the evolution of the effective optical depth. Their best values of $\Gamma_{\ion{H}{i}}$ are slightly smaller than our SDSS estimates, but again consistent within the error bars. In particular, their results also show no downturn of the photoionization rate towards large redshifts. On the other hand there is less agreement between our measurements and the latest estimates by by \citet{giguere08b}, which are  globally offset by $\sim 0.5$~dex towards low values with respect to the SDSS results (but again show no trace of a downturn). This discrepancy is presumably (at least partly) due to the already mentioned discrepant description of the optical depth evolution $\tau_\mathrm{eff}(z)$ (see Sect.~\ref{pe_sdss_txt:tau_evol}).

These estimates of the UVB can be compared to the semi-analytic prediction based on UV luminosity functions of quasars and star-forming galaxies \citep{haardt96,fardal98,haardt01}. As presented in Fig.~\ref{pe_sdss_fig:uvb_evol_comb}, the theoretical calculations done by \citet{haardt01} predict a significant decline of $\Gamma_{\ion{H}{i}}$ towards high redshifts, being consistent within the errors with both the results of \citet{bolton05} and our own from UVES. Since empirical constraints on both the quasar luminosity function and on the high-redshift stellar emissivity have been much improved in recent years, we present in the following a new calculation of the UVB.

\section{Origin of the metagalactic UV photons}\label{pe_sdss_txt:origin_uvb}
\subsection{Method}

The intensity of the background radiation field at a given frequency and redshift is derived from the cosmological radiative transfer equation
\begin{equation}
\left(\frac{\partial}{\partial t}-\nu H\frac{\partial}{\partial \nu}\right)J_\nu = -3HJ_\nu -c \alpha_\nu J_\nu+\frac{c}{4\pi}\epsilon_\nu
\end{equation}
 \citep{peebles93}, where $H(t)$ is the Hubble parameter, $c$ is the speed of light, and $\alpha_\nu $ and $\epsilon_\nu$ represent the absorption and emission coefficients, respectively (note that $\alpha_\nu$ has no connection with the quasar slope estimated in Sect.~\ref{pe_sdss_txt:comp_spec}). The average intensity of the radiation can be expressed, at an observed frequency $\nu_0$ and at a given redshift $z_0$, as
\begin{equation}
J_{\nu}\left(\nu_0,z_0\right) = \frac{1}{4\pi}\int_{z_0}^\infty \frac{(1+z_0)^3}{(1+z)^3}\ \epsilon_\nu(\nu,z)\ \mathrm{e}^{-\tau_\mathrm{Lyc}(\nu_0,z_0,z)}\ \frac{\mathrm{d}l}{\mathrm{d}z}\ \mathrm{d}z \label{pe_sdss_eq:jnu}
\end{equation}
with $\nu=\nu_0 (1+z)/(1+z_0)$, $\mathrm{d}l/\mathrm{d}z$ being the proper length element and $\tau_\mathrm{Lyc}$ is the Lyman continuum opacity, describing the radiation filtering due to intervening absorption.

The rapid increase of opacity along the LOS for $z\gtrsim 2$ leads to a significant reduction in the mean free path of ionizing photons, which as a consequence implies that only \emph{local} sources contribute significantly to the UV background intensity \citep{madau99,schirber03}. In this approximation, the solution of Eq.~\ref{pe_sdss_eq:jnu} is given by
\begin{equation}
J_{\nu}\left(\nu_0,z_0\right) \simeq \frac{\Delta l(\nu_0,z)\ \epsilon_\nu(\nu,z)}{4\pi}.
\end{equation}
The mean free path $\Delta l(\nu_0,z)$ is therefore related to the properties of the absorbers in the Ly$\alpha$ forest at redshift $z>2$ and is given at the Lyman limit frequency by
\begin{equation}
\Delta l(\nu_0,z) \simeq \frac{(\beta-1)\ c}{\Gamma(2-\beta)\ N_0 \sigma_\nu(\nu_0)^{\beta-1}}\ \frac{1}{(1+z)^{\gamma+1}\ H(z)}
\end{equation}
where  $\Gamma(2-\beta)$ in this case is the Gamma function and $(N_0,\beta,\gamma)$ describe the redshift and column density distributions of the absorbers,  typically well represented by 
\begin{equation}
\frac{\partial^2 N}{\partial z\ \partial N_\ion{H}{i}}=N_0\ N_\ion{H}{i}^{-\beta}\ (1+z)^{\gamma}.
\end{equation}
The major contribution to the mean opacity at the Lyman limit is driven by systems with optical depths around unity, or equivalently with column densities of the order of $N_\ion{H}{i}\simeq 10^{17.2}\ \mathrm{cm}^{-2}$. In this regime the properties of the absorbers in the Ly$\alpha$ forest are particularly difficult to infer and only few literature measurements exist \citep[e.g.][]{stengler95,kirkman97,peroux03}. For simplicity and due to the non-negligible contribution of lower column density systems to the Lyman continuum opacity \citep{madau99}, we approximated the column density distribution to a single power law with $\beta=1.5$. The absorber density evolution has been fixed to a power law as well with $(N_0,\gamma)=(1.4\times 10^7,2.45)$, following \citet{peroux03}. Deviations from these assumptions will only have a marginal influence on both the global and relative scaling of different contributions which, as we will show, does not affect our results.

\begin{figure}
\resizebox{\hsize}{!}{\includegraphics*[]{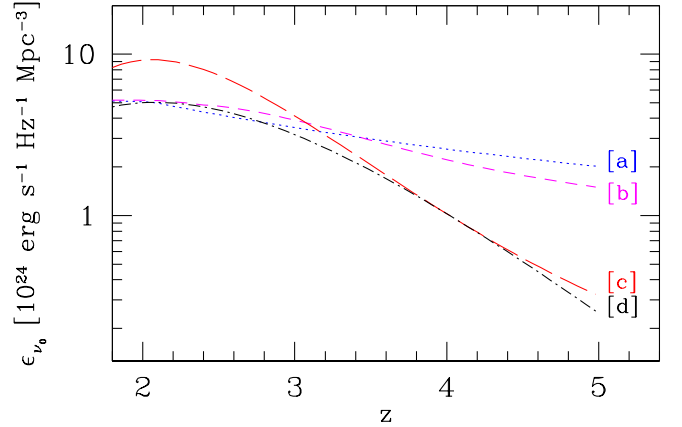}}
\caption{QSO emissivity $\epsilon_\nu(z)$ at the Lyman limit obtained from four different luminosity functions as described in the text. The different lines refers to [a] the \citet{bongiorno07} LDDE model,  [b] the modified LDDE model, [c] and [d] the \citet{wolf03} PLE and PDE model, respectively.}
\label{pe_sdss_fig:emissiv}
\end{figure}

Assuming that the UV background intensity has a power-law shape for wavelengths shorter than the Lyman limit ($J_\nu\propto\nu^{\alpha_\mathrm{b}}$), the UVB photoionization rate becomes
\begin{equation}
\Gamma_\mathrm{\ion{H}{i}}\left(z\right) \simeq \frac{\sigma_\nu(\nu_0)\ \Delta l(\nu_0,z)\ \epsilon_{\nu}(\nu_0,z)}{h\ (3-\alpha_\mathrm{b})}.\label{pe_sdss_eq:pir_hi}
\end{equation}
The spectral index of the UV background, $\alpha_\mathrm{b}$, is a poorly known quantity which in principle could be predicted assuming a specified contribution from quasars and galaxies \citep{haardt01}. In our calculation we fixed its value to $\alpha_\mathrm{b}=-1.4$ \citep{agafonova05}, but note that a different value would not change our conclusions as will be discussed in Sect.~\ref{pe_sdss_txt:pir_comp}. Equation~\ref{pe_sdss_eq:pir_hi} directly converts the integrated luminosity density at the Lyman limit, $\epsilon_\nu(\nu,z)$, into a photoionization rate and, assuming that only the quasar and star-forming galaxy populations contribute to $\Gamma_\mathrm{\ion{H}{i}}(z)$, it can be expressed as the sum of these two different contributions, thus
\begin{equation}
\Gamma_\mathrm{\ion{H}{i}}\left(z\right) \simeq \frac{\sigma_\nu(\nu_0)\ \Delta l(\nu_0,z)}{h\ (3-\alpha_\mathrm{b})}\ \left( \epsilon_{\nu,\mathrm{q}}(\nu_0,z) + \epsilon_{\nu,\mathrm{g}}(\nu_0,z)\right)\label{pe_sdss_eq:pir_hi2}
\end{equation}
where $\epsilon_{\nu,\mathrm{q}}(\nu_0,z)$ and $\epsilon_{\nu,\mathrm{g}}(\nu_0,z)$ represent the quasar and galaxy emissivities at the Lyman limit, respectively. In the following two sections we separately estimate and discuss these different contributors.

\begin{figure}
\resizebox{\hsize}{!}{\includegraphics*[]{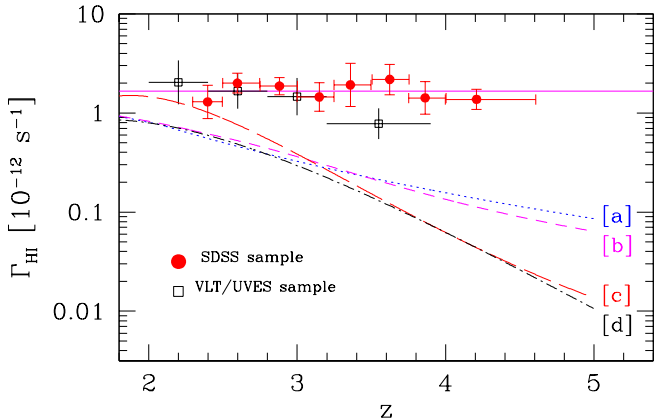}}
\caption{ QSO contributions to the UVB photoionization rate (labeled lines) as obtained from the different emissivities presented in Fig.~\ref{pe_sdss_fig:emissiv} in comparison to our estimates of the total photoionization rate from the SDSS (solid circles) and the UVES samples (empty squares). The solid line represent the assumed level for the total photoionization rate of $\log \Gamma_{\ion{H}{i}} = -11.78$. The different curves were inferred form [a] the \citet{bongiorno07} LDDE model,  [b] the modified LDDE model, [c] and [d] the \citet{wolf03} PLE and PDE model, respectively, of the quasar luminosity function.}
\label{pe_sdss_fig:gammaHI_qso}
\end{figure}

\subsection{Quasar contribution}\label{pe_sdss_txt:qso_contrib}

The integrated luminosity density, or emissivity, of quasars is defined as
\begin{equation}
\epsilon_{\nu,\mathrm{q}}(\nu,z) = \int_0^\infty \phi(L,\nu,z)\ L\ \mathrm{d}L \label{pe_sdss_eq:emissiv}
\end{equation}
where $\phi(L,\nu,z)$ is the quasar luminosity function (LF). In the literature, the quasar luminosity function has been estimated in a variety of rest frame bands and using different analytical descriptions. To convert from the observed passbands to the flux at the Lyman limit we approximated the quasar spectra energy distribution as a double power law ($f_\nu \propto \nu^\alpha$) with the following indices: $\alpha=-0.3$ at $2500-5000$ \AA\ \citep{madau99} and $-0.57$ at $228-2500$ \AA, as estimated in Sect.~\ref{pe_sdss_txt:comp_spec}. 

The quasar luminosity density has to be constrained through the results of quasar surveys. However, even though the number of known quasars has probably more than tenfolded in the last decade, the relevant range in the redshift-luminosity plane is still poorly sampled, mainly due to the fact that the luminosity density is dominated by the large number of faint QSOs which are hard to identify in large numbers. Instead of relying on a single global approximation, we looked into the results of different published surveys and their implications for the quasar luminosity density.

\noindent (i) \citet{wolf03} presented two different analytical fits to the COMBO-17 quasar luminosity function: A pure density evolution (PDE) and a pure luminosity evolution (PLE) model, both approximated by third order polynomials. The two derived emissivity laws predict a steep decrease towards high redshift with very similar slopes, but showing different peak contributions.

\noindent (ii) \citet{bongiorno07} used a combined sample of quasars from the VIMOS-VLT Deep Survey (VVDS) and SDSS, probing a larger luminosity interval compared to \citet{wolf03}. They employed a luminosity-dependent density evolution model (LDDE) with the limitation of a fixed bright end slope of the luminosity function. The resulting emissivity is considerably flatter than that of \citet{wolf03}.

\noindent (iii) \citet{richards06} estimated the luminosity function from the third SDSS data release (DR3). Their DR3 quasars are so bright that they were able to probe only the bright end slope of the luminosity function and they found significant evidence for a slope variation over redshift. For our purposes their luminosity function cannot be used since their analytical fits cannot be extrapolated to magnitudes fainter than $M_i \lesssim -26$. However we included the slope information into an additional LDDE model obtained from the \citet{bongiorno07} model, now allowing for a variable bright-end slope. The inferred emissivity decline at high redshift is slightly steeper, but still not as pronounced as in the two models by \citet{wolf03}. 

Figure~\ref{pe_sdss_fig:emissiv} presents a comparison of the emissivities estimated from the luminosity functions of \citet{wolf03,bongiorno07} and \citet{richards06}. We assume that these four analytic descriptions bracket the true evolution of the emissivity. Using Equation~\ref{pe_sdss_eq:pir_hi} we converted the luminosity density into an QSO contribution to $\Gamma_{\ion{H}{i}}$ which we denote as $\Gamma_{\ion{H}{i},\mathrm{q}}$.

\subsection{Contribution by star-forming galaxies}\label{pe_sdss_txt:star_contrib}

The second source population of the UV background to be considered are galaxies actively forming stars. The detection of high redshift galaxies is still extremely challenging and only a few deep surveys are available for this calculation. Among those reaching redshifts larger than $z=3$, we consider the results by \citet{tresse07} using the VVDS sample and by \citet{ouchi04} using the Subaru Deep Field (SDF) survey. In order to compute the galactic emissivity at the Lyman limit starting from the observed UV luminosity density we evaluated
\begin{equation}
\epsilon_{\nu,\mathrm{g}}(\nu,z)  = f_\mathrm{esc}\ g(\nu)\ \epsilon_\mathrm{UV}(\nu,z) .
\end{equation}
where $f_\mathrm{esc}$ is the escape fraction of UV photons from galaxies, $\epsilon_\mathrm{UV}(\nu,z)$ is the observed galactic UV emissivity per unit comoving volume, and $g(\nu)$ is the Lyman continuum frequency dependence of galaxies. We assumed that $g(\nu)\propto \nu^{-2}$ \citep[e.g.][]{inoue06} at all redshifts and a constant escape fraction of $10\%$. These quantities are highly uncertain and still controversially debated. However, we note that these assumptions affect the global scaling of the points and not their relative positions, unless one introduces a redshift-dependent escape fraction (which is possible, but adds another poorly constrained degree of freedom).  We then used Eq.~\ref{pe_sdss_eq:pir_hi} to convert the observed emissivities into a stellar contribution to $\Gamma_{\ion{H}{i}}$ which we denote as $\Gamma_{\ion{H}{i},\mathrm{g}}$. Notice that we refrained from attempting to construct a smooth interpolation (or even extrapolation) formula for $\epsilon_{\nu,\mathrm{g}}(\nu,z)$. We only evaluated the contribution by galaxies $\Gamma_{\ion{H}{i},\mathrm{g}}$ at the points given by the above mentioned surveys.

\begin{figure}
\resizebox{\hsize}{!}{\includegraphics*[]{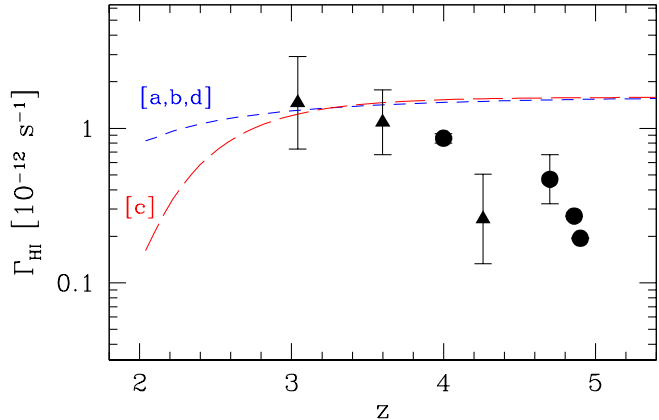}}
\caption{ Non-QSO contribution to the UV background photoionization rate. The different dashed lines represent the required photoionization rates which added to the QSO contribution lead to the observed constant  photoionization rate. The solid triangles and circles show $\Gamma_{\ion{H}{i}}$ estimated from the stellar emissivities of  \citet{tresse07} and \citet{ouchi04} respectively and assuming a constant escape fraction of 10\%. }
\label{pe_sdss_fig:gammaHI_star}
\end{figure}

\subsection{Comparison to the observed photoionization rate}\label{pe_sdss_txt:pir_comp}

We now consider how our measurements of the overall photoionization rate compare with the different contributions. Figure~\ref{pe_sdss_fig:gammaHI_qso} shows the different models of the quasar contribution to $\Gamma$ along with the estimates based on the proximity effect. The quasar contribution shows a peak around $z\simeq 2$  and strongly declines towards higher redshifts, however with slopes varying between the different models.

A number of factors make the QSO contribution to the photoionization rate uncertain. The analytical description of the quasar LF is by far the most prominent one, but also the assumed quasar SED, the slope of the UVB or the mean free path of ionizing photons come into play especially in the global scaling of $\Gamma_{\ion{H}{i}}$. Despite  the differences in the predicted $\Gamma_{\ion{H}{i},\mathrm{q}}(z)$, all models show a substantial decline in the quasar contribution to the global photoionization rate, implying that galaxies must provide a substantial additional contribution in order to maintain the IGM at the observed high ionization level.

In order to compare the UVB with the predictions from source counts, we subtracted the QSO contribution from the observed $\Gamma_{\ion{H}{i}}(z)$ obtained from the proximity effect analysis. For our SDSS measurements, the UVB is approximately constant at  $\Gamma_{\ion{H}{i}} = -11.78$. The residuals after subtracting the different luminosity densities are show in  Fig.~\ref{pe_sdss_fig:gammaHI_star}. In all cases $\Gamma_{\ion{H}{i},\mathrm{g}}(z)$ is dominating the global photoionization rate for redshifts larger than three, independently of the details of the assumed QSO contributions (models [a] to [d]). At smaller redshift the QSO contribution inferred from the PLE model by \citet{wolf03} (model [c]) is higher than that of the other models, and consequently the stellar contribution is smaller for this case. 

We also considered the extreme case of a UVB completely dominated by quasars at $z\sim2$, by rescaling all QSO emissivity models to our proximity effect measurement at $z\sim2$. This gave us the \emph{minimum} stellar contribution as a function of redshift. This is zero by design at  $z=2$, but increases rapidly and is in fact very similar to the model [c] in Fig.~\ref{pe_sdss_fig:gammaHI_star}. 

\begin{figure}
\resizebox{\hsize}{!}{\includegraphics*[]{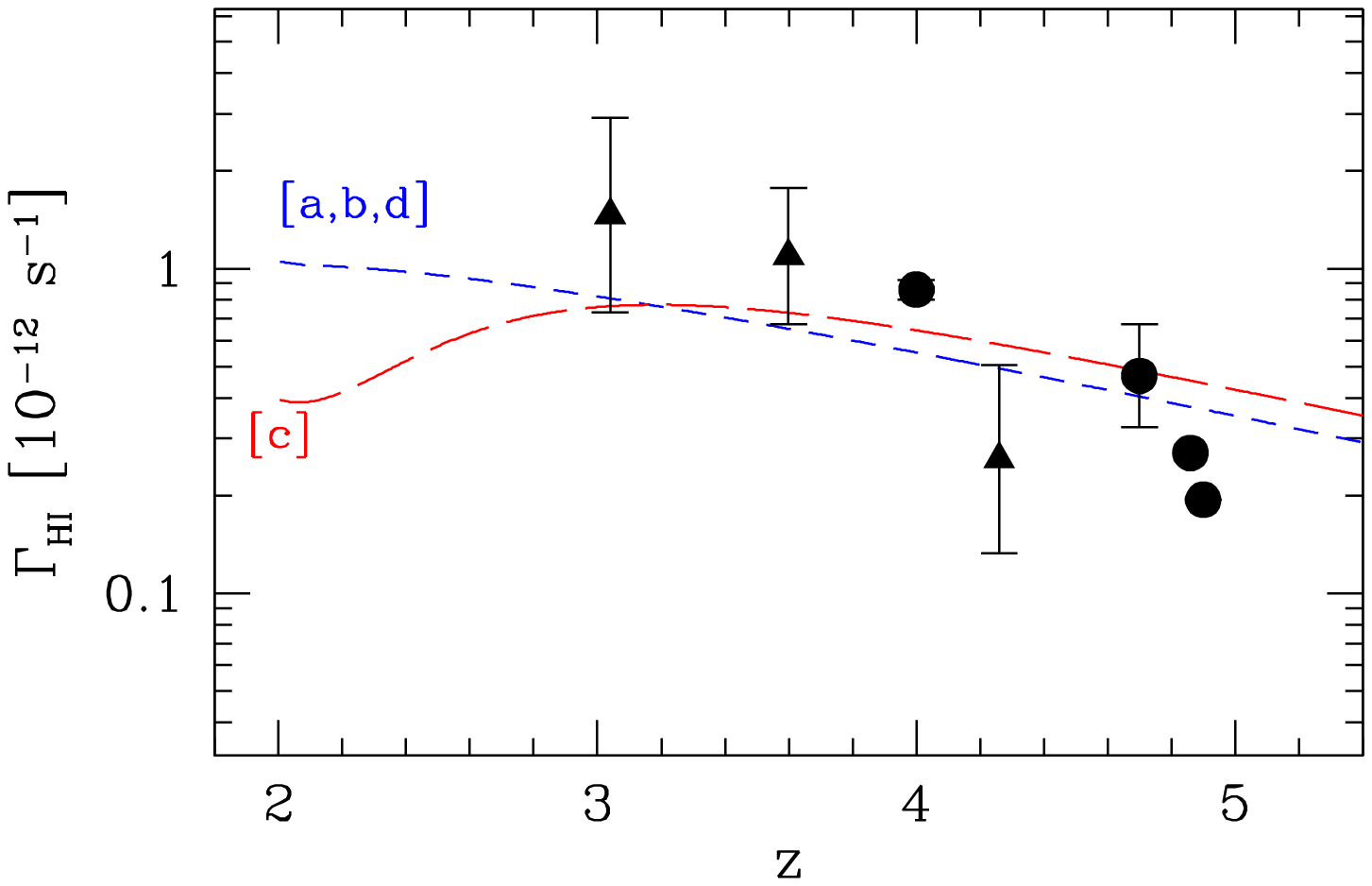}}
\caption{ Non-QSO contribution to the UV background photoionization rate as in Fig.~\ref{pe_sdss_fig:gammaHI_star}. The different dashed lines represent the required photoionization rates which added to the QSO contribution lead to a mildly decline in $\Gamma_{\ion{H}{i}}$ towards high redshift. The solid triangles and circles show $\Gamma_{\ion{H}{i}}$ estimated from the stellar emissivities of \citet{tresse07} and \citet{ouchi04} respectively and assuming a constant escape fraction of 10\%. }
\label{pe_sdss_fig:gammaHI_star_2}
\end{figure}

As the uncertainties on the global UVB photoionization rate are still considerable, we explored the impact of this uncertainty by alternatively adopting a slowly declining UVB as favored by our UVES results in Paper~II. We computed a linear fit to $\Gamma_{\ion{H}{i}}(z)$ and simply extrapolated this towards higher redshift. We then again subtracted the predicted QSO contribution and plotted the residuals in Fig.~\ref{pe_sdss_fig:gammaHI_star_2}. This  has a significant impact on the predicted stellar contribution. Recall, however, that the UVES data from Paper~II leave the UVB at $z\ga 4$ essentially unconstrained.

Fig.~\ref{pe_sdss_fig:gammaHI_star} and Fig.~\ref{pe_sdss_fig:gammaHI_star_2} constitute two independent assessments of the possible stellar contribution to the UVB. If $\Gamma_{\ion{H}{i}}$ at $z\simeq 4.2$ is as high as suggested by our highest-redshift SDSS data point, it would be fully consistent with the predictions based on the observed UV luminosity density of galaxies at $3 \la z \la 4$. Extrapolating our nearly flat UVB towards even higher redshifts would then result in an  increasing UVB deficit based on the observed galaxy surveys. However, that extrapolation is unsupported by the coverage of our data. Furthermore, some degree of deficit is also perfectly conceivable since high-redshift surveys as the SDF sample only the luminous part (considerably above $L^\star$) of the LBG population, and the number density of smaller star forming galaxies is still very poorly known.

If on the other hand the UVB should continue to decline towards higher $z$ as obtained by extrapolating the results from our UVES sample, the counts of star forming galaxies are in good agreement at $z > 4$, but almost too high in the better constrained range of $3 < z < 4$. We reiterate that these conclusions rest on the assumption of a constant UV escape fraction. By adjusting $f_\text{esc}$ and allowing for a variation with $z$, the stellar contribution to the UVB becomes essentially unconstrained.

\section{Conclusions}\label{pe_sdss_txt:concl}

Quantifying the metagalactic \ion{H}{i} photoionization rate at $z \ga 4$ remains a challenge. The three available methods all provide estimates that are not inconsistent with each other, but the uncertainties are still considerable. For example, the poorly constrained escape fraction of UV photons in galaxies makes a safe prediction of the UVB from population modeling virtually impossible at present. Likewise, the fundamentally very powerful method to match numerical simulations to the observed opacity evolution of the UVB has to rely on several theoretical assumptions that influence the outcome of the simulations. The proximity effect is arguably the most direct way to measure a quantity that can be translated into an estimate of $\Gamma_{\ion{H}{i}}$. The fundamental issue here is the question whether large-scale \ion{H}{i} overdensities around quasars can weaken the proximity effect and thus bias the UVB towards high values. Our SDSS sample provides clear evidence for the presence of such overdensities, but our analysis also suggests that the fraction of QSOs heavily affected by overdensities is small. Certainly our data point at $z\simeq 4.2$ is very robust and unlikely to be affected by any significant overdensity bias.

Compared to previous analyses of the proximity effect in quasar spectra, the main improvement of the present study lies the size of the dataset. It constitutes an increase by a factor of $\sim 20$ compared to the hitherto largest sample by \citet{scott00}, at similar spectral resolution, and by a factor of 50 compared to our recent evaluation of 40 high-resolution spectra from VLT/UVES (Paper~II). At redshifts $z>4$, our new SDSS dataset contains 34 objects, whereas previous investigations of the proximity effect at such high redshifts were always based on individual objects
\citep{williger94,lu96,savaglio97}. 

We have also introduced an automated continuum determination and correction algorithm, which is of fundamental importance in quantifying the inevitable effects of continuum misplacement. The quality of this correction is impressively demonstrated by the fact that our inferred global evolution of the Ly$\alpha$ forest optical depth is very close to the results of most recent studies using high spectral resolution. We thus conclude that the moderate spectral resolution of the SDSS data has been well accounted for and is not an issue.

In continuation of our previous work on the proximity effect (Papers~I and II) we have searched for the proximity effect signature not only in the combined sample, but in individual quasar spectra. We detected the effect in 98~\% of the objects, with the remaining 2~\% all clearly affected by strong associated absorption. With a sample of this size we could construct the `proximity effect strength distribution' (PESD) in several redshift bins, with 30--600 QSOs contributing to each bin. The observed proximity effect strengths display a considerable dispersion and a pronounced asymmetry. Both the dispersion and the skewness of the PESD decrease with increasing redshift, just as predicted from a simple model of randomly distributed absorption lines following the global line density redshift evolution. While it is clear that quasar absorption lines are not exactly randomly distributed but follow the cosmic web, the distribution of pixel optical depths can be reasonably well described as a random process. In Paper~II we predicted a very strong change of the PESD width with redshift, but owing to a sample size of only 40 objects we could not demonstrate the reality of the effect. The SDSS data show this trend very clearly. 

The fact that the PESD at higher redshift becomes increasingly narrow has profound implications for the accuracy of measuring the UV background. While any QSO at $z \sim 2$ may show a proximity effect strength that deviates by up to 2 orders of magnitude from the expectation value, the dispersion is much lower at $z\ga 4$, and relatively small samples, even individual objects can give meaningful constraints on the UVB. This is impressively confirmed by the subset of 34 QSOs at $z>4$ in our sample.

While our measurement of the metagalactic \ion{H}{i} photoionization rate at $z\sim 4$ has yielded a rather high value, it is nevertheless fully consistent with most other recent estimates, certainly within 1--$2\sigma$ deviations. There is thus a clear trend that the UVB is almost constant at \mbox{$\log \Gamma_{\ion{H}{i}} = -11.78\pm 0.07$~s$^{-1}$}\ between $z\simeq 2$ and $z\simeq 4$, at best it decreases very mildly. It will be a most challenging task in the future to push the current limit of $z\simeq 4$ towards higher redshifts and see whether the UVB continues to be flat, or whether a downturn will be detected.

\begin{acknowledgements}
We would like to thank the Astrophysical Research Consortium (ARC) for making the Sloan Digital Sky Survey (SDSS) data archive publicly available. A.D. and L.W. acknowledge support by the Deutsche Forschungsgemeinschaft under Wi~1369/21-1.
\end{acknowledgements}


\bibliography{bibliography}

\end{document}